\documentclass[9pt]{extarticle}
\usepackage[T1]{fontenc}
\usepackage{spconf,amsmath,graphicx}
\usepackage{cite}
\usepackage{amssymb,amsfonts}
\usepackage{graphicx,color}
\usepackage{textcomp}
\usepackage{xcolor}
\usepackage{hyperref}
\usepackage{algorithm}
\usepackage{amsmath,bm,mathtools,mathrsfs,mathabx}
\usepackage{multicol}
\usepackage{amsthm}
\usepackage{float}
\usepackage{comment}

\usepackage{cleveref}
\usepackage{enumitem}
\usepackage{outlines}
\usepackage{algpseudocode}
\usepackage[skip=0pt]{caption}
\usepackage[skip=0pt]{subcaption}
\AtBeginDocument{\definecolor{tmlcncolor}{cmyk}{0.93,0.59,0.15,0.02}\definecolor{NavyBlue}{RGB}{0,86,125}}
\newcommand\scalemath[2]{\scalebox{#1}{\mbox{\ensuremath{\displaystyle #2}}}}

\newtheorem{theorem}{Theorem}

\newtheorem{assumption}{Assumption}

\newtheorem{definition}{Definition}

\newcommand*{\hermtr}{{\mathsf{H}}}

\usepackage{textcomp}

\usepackage{bm}
\makeatletter
\AtBeginDocument{\DeclareMathVersion{bold}
\SetSymbolFont{operators}{bold}{T1}{times}{b}{n}
\SetMathAlphabet{\mathrm}{bold}{T1}{times}{b}{n}
\SetMathAlphabet{\mathit}{bold}{T1}{times}{b}{it}
\SetMathAlphabet{\mathbf}{bold}{T1}{times}{b}{n}
\SetMathAlphabet{\mathtt}{bold}{OT1}{pcr}{b}{n}
\SetSymbolFont{symbols}{bold}{OMS}{cmsy}{b}{n}
\renewcommand\boldmath{\@nomath\boldmath\mathversion{bold}}}

\def\L{{\cal L}}

\title{Data-Driven Two-Stage IRS-Aided Sumrate Maximization with Inexact Precoding \vspace{-6pt}}
\name{Hassaan Hashmi$^\dagger$, Spyridon Pougkakiotis$^\ddagger$, and Dionysis Kalogerias$^\dagger$
\vspace{-10pt}}
\address{$^\dagger$Department of Electrical and Computer Engineering, Yale University, New Haven, USA\\
$^\ddagger$Department of Mathematics, King’s College London, London, England, UK\\
 \small{\tt \href{mailto:hassaan.hashmi@yale.edu}{hassaan.hashmi@yale.edu}, \href{mailto:spyridon.pougkakiotis@kcl.ac.uk}{spyridon.pougkakiotis@kcl.ac.uk}, \href{mailto:dionysis.kalogerias@yale.edu}{dionysis.kalogerias@yale.edu}}
\vspace{-14pt} \thanks{This work is supported by the US NSF under Grants 2242215 and 2431860}.}
\begin{document}
%
\maketitle
\begin{abstract}
\vspace{-0.5pt}
We propose \textit{iZoSGA}, a data-driven learning algorithm for joint passive long-term intelligent reflective surface (IRS)-aided beamforming \textit{and} active short-term precoding in wireless networks. iZoSGA is based on a zeroth-order stochastic quasigradient ascent methodology designed for tackling two-stage nonconvex stochastic programs with continuous uncertainty and objective functions with ``black-box'' terms, and where second-stage optimization is inexact. As such, iZoSGA utilizes \textit{inexact precoding oracles}, enabling practical implementation when short-term (e.g., WMMSE-based) beamforming is solved approximately. The proposed method is agnostic to channel models or statistics, and applies to arbitrary IRS/network configurations. We prove non-asymptotic convergence of iZoSGA to a neighborhood of a stationary solution of the \textit{original exact problem} under minimal assumptions. Our numerics confirm the efficacy iZoSGA in several ``inexact regimes'', enabling passive yet fully effective IRS operation in diverse and realistic IRS-aided scenarios.

\end{abstract}
\vspace{-5pt}

\vspace{-10pt}
\section{Introduction}\label{sec:intro}
\vspace{-8.5pt}
\par
\begin{figure}[!t]
  \centering
  \centerline{\includegraphics[width=2.5in]{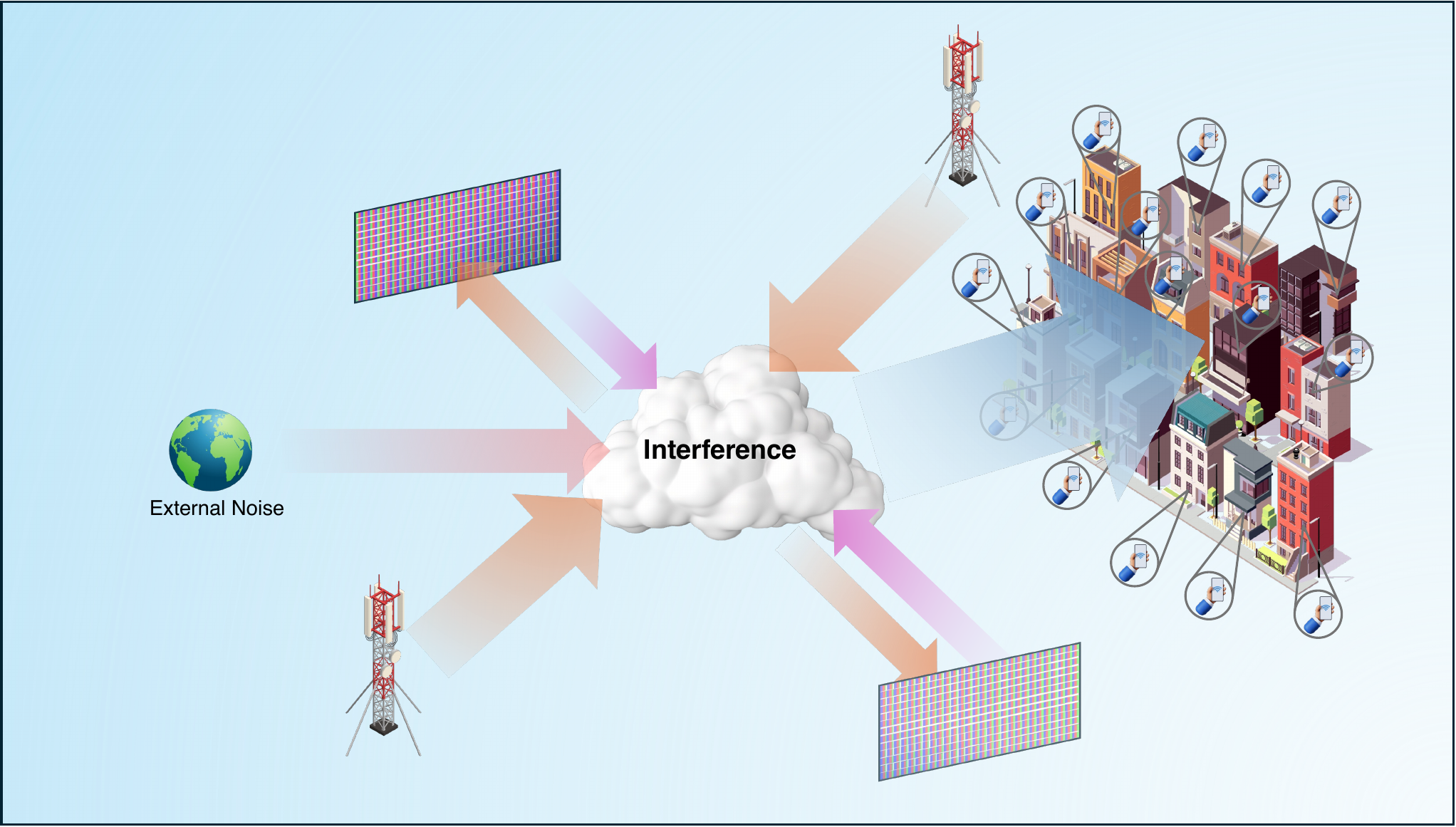}}
\vspace{4pt}
\caption{Depiction of a realistic IRS-aided Wireless Network.}
\label{fig:sketch}
\vspace{-16pt}
\end{figure}

 The growing need for improving  quality of service (QoS) for efficient and reliable wireless communication necessitates perpetual investigation into novel technologies\cite{mimo:larsson2014massive, udn:kamel2016ultradense, udn:gotsis2016ultradense, 5g:andrews2014will, 5g:shafi20175g}. One such technology is Intelligent Reflecting surfaces (IRSs, aka RISs), i.e., planar metasurfaces comprised of (a large number of) passive scattering elements with tunable phase shifts and amplitudes, capable of being dynamically adjusted to manipulate the propagation of incident waves. Even in absence of line-of-sight (LoS), IRSs function as \textit{channel shapers} by creating alternative paths between access points (APs) and receivers; see Figure \ref{fig:sketch}.

\par The goal of deploying IRSs in a wireless network is to optimally tune their phase-shifting elements concurrently with other resources, such as AP precoders, in order to enhance QoS. In this work, we consider the MISO downlink scenario and optimize the \textit{weighted sumrate utility}, but other metrics can also be considered. 

\par Recently, the potentially far-reaching benefits of deploying \textit{passive} IRSs to enhance standard beamforming schemes have led to the emergence of a rich literature on methods for IRS-aided optimal beamforming. These methods can be categorized in terms to two general system models. The \textit{first system model} treats IRS elements as short-term (reactive) beamformers, tuned simultaneously with transmit precoding using instantaneous cascaded  channel state information (CSI), thus relying on detailed channel/system models, resulting in specialized solvers; see, e.g., \cite{reactive:wu2019intelligent, reactive:wu2019towards, reactive:wang2020intelligent, reactive:wu2020joint,sco:yang2021,10123709}. This reliance on continuous, fine-grained instantaneous CSI renders such methods operationally unrealistic and computationally prohibitive in practical scenarios. While deep learning methods \cite{ml:taha2021,ml:yang2021, drl:mismar2019dqn,drl:huang2020ddpg} ease some limitations, using deep neural networks can easily lead to increased problem complexity, opaqueness, poor generalization, and/or overfitting.

\par The \textit{second system model}, which is adopted here as well, optimizes IRS elements over the long-term timeframe while also applying standard short-term (downstream) transmit precoding. Although much more realistic from a practical perspective, this model results in challenging \textit{nonconvex two-stage stochastic programming} problems. Up until recently, \textit{Stochastic Successive Convex Approximation} (SSCA) techniques \cite{SSCO_1, SSCO_2} were considered to be state-of-the-art (SotA) for optimally tuning IRSs for this system model. However, SSCA-based techniques are dependent on explicit channel and network topology models, accurate cascaded (statistical) CSI estimation, and require the formulation of convex surrogates of the original (highly nonconvex) problems \cite{sco:zhao2020tts}.

To successfully mitigate all aforementioned limitations, a novel, fully model-free and data-driven zeroth-order algorithm, termed \textit{ZoSGA}, was proposed recently \cite{IEEE_tran_Hashmietal}. It was shown, both theoretically and empirically, that ZoSGA (ergodically) converges in the vicinity of some critical point of the given (nonconvex) two-stage stochastic program, substantially outperforming previous SotA SSCA-based techniques. However, in \cite{IEEE_tran_Hashmietal} it was critically assumed that ZoSGA has access to a \textit{perfect oracle} for short-term precoding (while usually implemented via WMMSE \cite{wmmseShi2011}). While ZoSGA is vastly more versatile and effective than its rivals (i.e., the previous SotA), such an assumption creates a nontrivial gap between theory and practice, especially related to actual potential deployability of ZoSGA \cite{IEEE_tran_Hashmietal}.

\par This work bridges this gap by introducing and analyzing a generalized form of the algorithm in \cite{IEEE_tran_Hashmietal}, in the context of sumrate maximization. This new algorithm, termed iZoSGA, relies on \textit{inexact precoding oracles}, which may provide \textit{arbitrarily sub-optimal} short-term beamformers. Our contributions are summarized as follows:
\vspace{-5pt}
\begin{itemize}[leftmargin=*]
    \item We provide a non-asymptotic analysis of iZoSGA. We show that oracle inexactness contributes an additional \textit{iterate-averaged} error term to the final convergence rate, plausibly  establishing \textit{stability of iZoSGA under precoding inexactness}. 
    \vspace{-7pt}
    \item We elucidate how oracle errors can be controlled with minimal assumptions on the IRS-aided two-stage beamforming problem.
    \vspace{-7pt}
    \item We empirically demonstrate how our analysis can be used for the \textit{development of practical algorithmic variants of iZoSGA}, for tackling realistic IRS-aided optimal beamforming problems. 
\end{itemize}
\vspace{-5pt}
\textit{Note:} Proofs are omitted, to be included in an extended journal submission. Code is available at: \href{https://github.com/hassaanhashmi/izosga}{github.com/hassaanhashmi/izosga}.



\section{Problem Preliminaries}
\label{sec:prob_formulation}
\vspace{-7pt}
\par Following the \textit{second system model} as in \cite{IEEE_tran_Hashmietal}, we assume a MISO downlink network setting with $K$ receivers actively serviced by an AP with $M$ antennas, and consider the well known twice continuously (real) differentiable weighted sumrate utility defined as 
\begin{equation}\nonumber
\scalemath{0.97}{{F}(\bm{W},\bm{H}(\bm{\theta},\omega)) \triangleq \sum_{k=1}^K \alpha_k \log_2\left(1 + \textnormal{SINR}_k\left(\bm{W},\bm{h}_k\left(\bm{\theta},\omega  \right)\right)\right)},
\end{equation}
where each $\alpha_{k}>0$ is a weighting scalar for receiver $k \in \mathbb{N}^+_{K}$, and where the $\textnormal{SINR}_k\left(\cdot,\cdot\right)$ term captures the per-receiver signal-to-noise plus interference ratio expressed as 
\begin{align}\nonumber
\begin{aligned} \label{eq:2}
\text{SINR}_k(\bm{W},\bm{h}_k(\bm{\theta}, \omega )) \triangleq \frac{\left|\bm{h}_k^\hermtr(\bm{\theta}, \omega)\bm{w}_k\right|^2}{{\sum}_{{j \in \mathbb{N}_{K}^+ \setminus k}}\left|\bm{h}^\hermtr_k(\bm{\theta}, \omega)\bm{w}_j\right|^2 + \sigma^2_k}.
\end{aligned}
\end{align}
In the above expressions, transmit precoding vectors are denoted by $\bm{W} {=} \mathrm{vec}(\begin{bmatrix} \bm{w}_1 & \bm{w}_2 & {\cdots} & \bm{w}_K \end{bmatrix})\ {\in}\ \mathbb{C}^{M_U\triangleq (M \times K)}$, and the \textit{effective} channel matrix by $\bm{H}(\bm{\theta},\omega) \triangleq \mathrm{vec}(\begin{bmatrix} \bm{h}_1 \ldots \bm{h}_K \end{bmatrix})\ {\in}\ \mathbb{C}^{M_U}$, where $\bm{\theta}$ denotes tunable IRS parameters and $\omega \in \Omega$ denotes a \textit{state of nature}, representing unobservable random propagation patterns generating effective CSI (for each choice of fixed IRS parameters $\bm{\theta}$). Lastly, $\sigma^2_{k}$ is the receiver-specific noise variance. 

\par We consider a compact set $\mathcal{W}$ of feasible beamformers informed by a power budget of $P>0$ at the AP, constraining precoders $\bm{W}$ such that $\|\bm{W}\|_F^2 \leq P$, where $\|\cdot\|_F^2$ denotes the Frobenius norm. Then, following our prior work in \cite{IEEE_tran_Hashmietal}, we are interested in the two-stage (short/long-term) stochastic beamforming problem of the form
\begin{equation} \label{eqn: two-stage problem} \tag{2SP}
\begin{split}
    \boxed{
    \max_{\boldsymbol{\theta} \in \Theta} f(\bm{\theta}) \triangleq \mathbb{E}\left\{\max_{\boldsymbol{W} \in \mathcal{W}}{F}\left(\bm{W},\boldsymbol{H}(\boldsymbol{\theta},\omega)\right) \right\},}
    \end{split}
\end{equation}
where $\Theta \in \mathbb{R}^S$ is a convex and compact set of feasible IRS parameters $\bm{\theta}$, i.e., corresponding to amplitudes and phases; note that \textit{no} unit-modulus constraints are necessary in our development; see also our previous work in \cite{IEEE_tran_Hashmietal} (and our numerical results later in Section \ref{sec:simulations}). For any $\omega \in \Omega$ and any $\bm{\theta} \in \Theta$, the deterministic \textit{second-stage} precoding problem is expressed as 
\begin{equation} \label{eqn: second-stage problem} \tag{SSP}
\max_{\bm{W} \in \mathcal{W}} \,\, \left\{G(\bm{W},\bm{\theta},\omega)   \triangleq F\left(\bm{W},\bm{H}(\bm{\theta},\omega)\right)\right\}.
\end{equation}
Hereafter, the \textit{IRS-shaped} effective channel $\bm{H}(\bm{\theta},\omega)$ is considered a \textit{black-box with unknown dynamics}, but we assume that it is sufficiently regular (cf. Assumption \ref{assumption: two-stage problem} later in Section \ref{sec:conv_anal}). Under this model-free setting, we rely on two-point stochastic evaluation of $\bm{H}(\cdot,\omega)$ \cite{IEEE_tran_Hashmietal} to estimate its gradients as
\begin{equation} \nonumber
\begin{split}
\hspace{-2pt}\nabla_{\bm{\theta}} \bm{H}_\mu(\bm{\theta},\omega) \hspace{-2pt}\triangleq\hspace{-1.5pt}
&\ \mathbb{E}\hspace{-2pt}\left\{\frac{\bm{H}\left(\bm{\theta} + \mu \bm{U},\omega\right) - \bm{H}\left(\bm{\theta}-\mu \bm{U},\omega\right)}{2\mu} \bm{U}^\top \bigg\vert \ \omega \right\}^\top,
\end{split}
\end{equation}
where $\bm{U}\in \mathcal{N}(\bm{0},\bm{1})$, and $\mu>0$ is a given smoothing parameter. Then, the compound channel zeroth-order approximation of the gradient of ${F}(\bm{W},\bm{H}(\bm{\theta},\omega))$ relative to $\boldsymbol{\theta}$ is given by \cite[Lemma 1]{IEEE_tran_Hashmietal}
\begin{align} \label{eqn: ZO gradient of the compositional function}
  & \nabla_{\bm{\theta}}^{\mu} F\left(\bm{W},\bm{H}(\bm{\theta},\omega)\right) \\
  &\quad \triangleq \ 2 \nabla_{\bm{\theta}} \bm{H}_{\mu}^R(\bm{\theta},\omega)\left(\Re\left(\frac{\partial^{\circ}}{\partial \bm{z}} F(\bm{W},\bm{z}) \big\vert_{\bm{z} = \bm{H}(\bm{\theta},\omega)} \right)\right)^\top \nonumber \\ & \qquad + 2\nabla_{\bm{\theta}}  \bm{H}_{\mu}^I(\bm{\theta},\omega)\left(\Re\left(j\frac{\partial^{\circ}}{\partial \bm{z}} F(\bm{W},\bm{z}) \big\vert_{\bm{z} = \bm{H}(\bm{\theta},\omega)} \right)\right)^\top, \nonumber
\end{align}
\noindent where $\frac{\partial^{\circ}}{\partial \bm{z}}$ denotes the Wirtinger co-gradient, and superscript ``R'' (resp. ``I'') denotes real (resp. imaginary) part. It should be noted here that we avoid explicit channel modeling assumptions that introduce and accumulate errors, particularly in large-scale systems.

\section{Inexact Precoding Oracle}
\label{sec:izosga}
\vspace{-7pt}

To expand on the role of $\bm{W}$ in the gradient expression in \eqref{eqn: ZO gradient of the compositional function}, we formalize the notion of \textit{inexactness} in the context of the inner problem. To this end, we define an \textit{inexact oracle} as a black-box that returns points that are ``close'' to an arbitrary optimal solution of \eqref{eqn: second-stage problem}.

\begin{definition} \label{definition: oracle assumption}
For any $\bm{\theta} \in \Theta$ and a.e. $\omega \in \Omega$, an \emph{inexact precoding oracle} returns an approximate precoder $\widetilde{\bm{W}} \in \mathcal{W}$ to \textnormal{\eqref{eqn: second-stage problem}}, such that
\begin{equation} \label{eqn: solution distance error} \mathrm{dist}\Big(\widetilde{\bm{W}},\arg\max_{\bm{W} \in \mathcal{W}} F(\bm{W},\bm{H}(\bm{\theta},\omega))\Big) \leq \varepsilon(\bm{\theta},\omega),
\end{equation}
\noindent with $\varepsilon(\cdot,\cdot)$ some measurable random error function.
\end{definition}
\par From  Definition \ref{definition: oracle assumption}, we have that the distance between inexact oracle output $\widetilde{\bm{W}}$ and some $\bm{W}^* \in \arg\max_{\bm{W} \in \mathcal{W}} F(\bm{W},\bm{H}(\bm{\theta},\omega))$ is such that $\|\widetilde{\bm{W}} - \bm{W}^*\| \leq \varepsilon(\bm{\theta},\omega)$. Here, we have only imposed measurability on the random error function $\varepsilon(\cdot,\cdot)$. This abstraction allows us to incorporate various inexact oracles in this generic framework. In this work, we will employ the well-documented WMMSE \cite{wmmseShi2011} as the inexact oracle for solving the weighted sumrate maximization problem. Then, $\varepsilon$ can be completely characterized by the accuracy and convergence behavior of the sequence of iterates generated by WMMSE, as we shall see in Section \ref{sec: compatibility of assumptions}.

\vspace{-4pt}
\section{A New Method for The Outer Problem}
\label{sec:izosga}
\vspace{-7pt}


We employ inexact precoding oracles, characterized by error $\varepsilon$ (Definition \ref{definition: oracle assumption}), to develop a model-free projected stochastic quasi-gradient method for the two-stage beamforming problem. Given i.i.d. samples of $\omega$, we evaluate \eqref{eqn: ZO gradient of the compositional function} at each $(\bm{\theta},\bm{W}) \in \Theta \times \mathcal{W}$ (for a.e.\ $\omega$), producing the zeroth-order gradient approximation
\begin{align} \label{eqn: ZO gradient sample of the compositional function}
 & \bm{D}_{\mu}(\bm{\theta},\omega,\bm{U};\bm{W})  \triangleq \bm{\Delta}_{\mu}^R \left(\Re\left(\frac{\partial^{\circ}}{\partial \bm{z}} F(\bm{W},\bm{z}) \bigg\vert_{\bm{z} = \bm{H}(\bm{\theta},\omega)} \right)\right)^\top \nonumber\\ & \qquad   + \bm{\Delta}_{\mu}^I \left(\Re\left(j\frac{\partial^{\circ}}{\partial \bm{z}} F(\bm{W},\bm{z}) \bigg\vert_{\bm{z} = \bm{H}(\bm{\theta},\omega)} \right)\right)^\top,
\end{align}
where, $\bm{\Delta}_{\mu}^R$ and $ \bm{\Delta}_{\mu}^I$ are respectively sample-based evaluations of the expected gradient approximations in \eqref{eqn: ZO gradient of the compositional function}. The induced overhead of two additional channel estimations is implicitly (and reasonably) assumed to be within the coherence time of $\bm{H}$.

\par Algorithm \ref{Algorithm: iZoSGA-WMMSE} computes a projected quasi-gradient by evaluating the zeroth-order gradient sample at \(\bm{W}{=}\widetilde{\bm{W}}\) given by the WMMSE algorithm, in line with Definition \ref{definition: oracle assumption}. This constitutes a rough Gaussian smoothing–based approximation of the second-stage gradient, derived using an inexact oracle rather than a proper surrogate. In total, each iteration requires exactly three channel estimations, one for short-term precoding and two more for system probes required for gradient sampling. Further, any standard channel estimation method suffices, as it is independent of the presence of  IRSs.

\begin{algorithm}[!t]
\caption{iZoSGA with WMMSE}
    \label{Algorithm: iZoSGA-WMMSE}
\begin{algorithmic}
\State \textbf{Input:}  $\bm{\theta}_0 \in \Theta$, $\eta > 0$, $\mu > 0$, $\{T_{\text{WMMSE},t}\}_{t=0}^T$, and $T > 0$.
\For {($t = 0,1,2,\ldots, T$)}
\State Sample (i.i.d.) $\omega_t \in \Omega$, $\bm{U}_t \sim \mathcal{N}\left(\bm{0},\bm{I}\right)$.
\State Call WMMSE for $T_{\text{WMMSE},t}$ iterations to obtain $\widetilde{\bm{W}}_t$.
\State Set $\bm{D}_{\mu,t} \equiv \bm{D}_{\mu}\left(\bm{\theta}_t,\omega_t,\bm{U}_t;\widetilde{\bm{W}}_t\right)$ as in \eqref{eqn: ZO gradient sample of the compositional function}.
\State  $\bm{\theta}_{t+1} = \textnormal{\textbf{proj}}_{\Theta}\left(\bm{\theta}_t + \eta \bm{D}_{\mu,t}\right). $
\EndFor
\State Sample $t^* \in \{0,\ldots,T\}$ according to $\mathbb{P}(t^* = t) = \frac{1}{T+1}$.
\State \Return $\bm{\theta}_{t^*}$.
\end{algorithmic}
\end{algorithm}
\subsection{Controlling the Inexact Precoding Oracle Error}
\label{sec: compatibility of assumptions}

In light of nonconcavity of \eqref{eqn: second-stage problem}, let us explain under what conditions one can effectively control the magnitude of the oracle error function $\varepsilon(\bm{\theta},\omega)$ (cf. Definition \ref{definition: oracle assumption}). In our previous work \cite[Appendix B]{IEEE_tran_Hashmietal}, it was shown that the function $\max_{\bm{W}\in\mathcal{W}}F(\bm{W},\cdot) - F(\cdot,\cdot)$ is \textit{sub-analytic}. It then follows that this function satisfies the \emph{\L{}ojasiewicz inequality} with uniform exponent \cite{LojasiewiczInequality}. Thus, there exists $\xi>0$ and a positive sub-analytic function $C(\bm{H}(\cdot,\cdot))$, such that for all $\omega\in\Omega$, $\bm{\theta}\in\Theta$, and all $\bm{W}\in\mathcal{W}$,
    \begin{align} \label{eqn: Lojasiewicz}
        &\textnormal{dist}\Big(\bm{W}, \arg\max_{\bm{W} \in \mathcal{W}} F\left(\bm{W},\bm{H}(\bm{\theta},\omega)\right) \hspace{-2pt}
        \Big) 
        \\ & \ \leq C(\bm{H}) \Big(\max_{\bm{W} \in \mathcal{W}} F(\bm{W},\bm{H}(\bm{\theta},\omega)) - F(\bm{W},\bm{H}(\bm{\theta},\omega))\Big)^{\xi}. \nonumber
    \end{align}
\noindent Under the minor assumption that there exists a uniform constant $C_{\bm{H}} \geq C(\bm{H}(\bm{\theta},\omega))$ for any $\bm{\theta}\in\Theta$ and any $\omega\in\Omega$, it then follows that the magnitude error in Definition \ref{definition: oracle assumption} can be directly controlled.
\par Indeed, as \eqref{eqn: second-stage problem} is nonconcave, we can only find an approximate solution $\widetilde{\bm{W}}$, such that
\begin{equation} \label{eqn: natural error criterion}
\max_{\bm{W} \in \mathcal{W}} F(\bm{W},\bm{H}(\bm{\theta},\omega)) - F\big(\widetilde{\bm{W}},\bm{H}(\bm{\theta},\omega)\big) \leq \widetilde{\varepsilon}(\bm{\theta},\omega),
\end{equation}
where the magnitude of $\widetilde{\varepsilon}(\bm{\theta},\omega)$ can be reasonably controlled (in our case by adjusting the number of WMMSE iterations utilized to obtain $\widetilde{\bm{W}}$). From \eqref{eqn: Lojasiewicz}, it follows that, for all $\bm{\theta} \in \Theta$ and a.e. $\omega \in \Omega$,
\begin{equation} \label{eqn: final oracle error bound} \big\|\widetilde{\bm{W}} - \bm{W}^* \big\| \leq C_{\bm{H}} \widetilde{\varepsilon}(\bm{\theta},\omega)^{\xi}, \end{equation}
where  $\bm{W}^* \in \arg\max_{\bm{W} \in \mathcal{W}} F(\bm{W},\bm{H}(\bm{\theta},\omega))$. Thus, the magnitude of the oracle error of Definition \ref{definition: oracle assumption} can be controlled by enforcing \eqref{eqn: natural error criterion}. This supports the practical efficiency of Algorithm \ref{Algorithm: iZoSGA-WMMSE},  which relies on approximate WMMSE-based \cite{wmmseShi2011} solutions to \eqref{eqn: second-stage problem}.

\vspace{-4pt}
\subsection{Convergence Analysis}\label{sec:conv_anal}

We briefly outline the convergence properties of Algorithm \ref{Algorithm: iZoSGA-WMMSE}. We write the  objective of \eqref{eqn: two-stage problem}  as $\phi(\bm{\theta}) \triangleq -f(\bm{\theta}) + \delta_{\Theta}(\bm{\theta})$, where $\delta_{\Theta}(\bm{\theta})$ is the indicator over the convex set $\Theta$, and introduce the \textit{Moreau envelope} associated with $\phi$, with parameter $\lambda > 0$:
\[ \phi^{1/\lambda}(\bm{u}) \triangleq \min_{\bm{\theta} \in \mathbb{R}^S} \left\{\phi(\bm{\theta}) + \frac{\lambda}{2}\|\bm{u}-\bm{\theta}\|^2 \right\}.\]
Since $\phi$ is $\rho$-weakly convex (as shown in \cite{IEEE_tran_Hashmietal}), $\phi^{1/\lambda}$ is smooth for $\lambda > \rho$ and its gradient norm serves as a near-stationarity measure. In that sense, an $\epsilon$-stationary point of the Moreau envelope implies proximity to a near-stationary point (in the Clarke sense) of the original nonsmooth problem  (e.g., see \cite[Section 2]{SIAMOpt:Davis}). Before establishing convergence, we state our basic assumptions below.

\begin{figure}
  \centering
  \centerline{\includegraphics[width=2.7in]{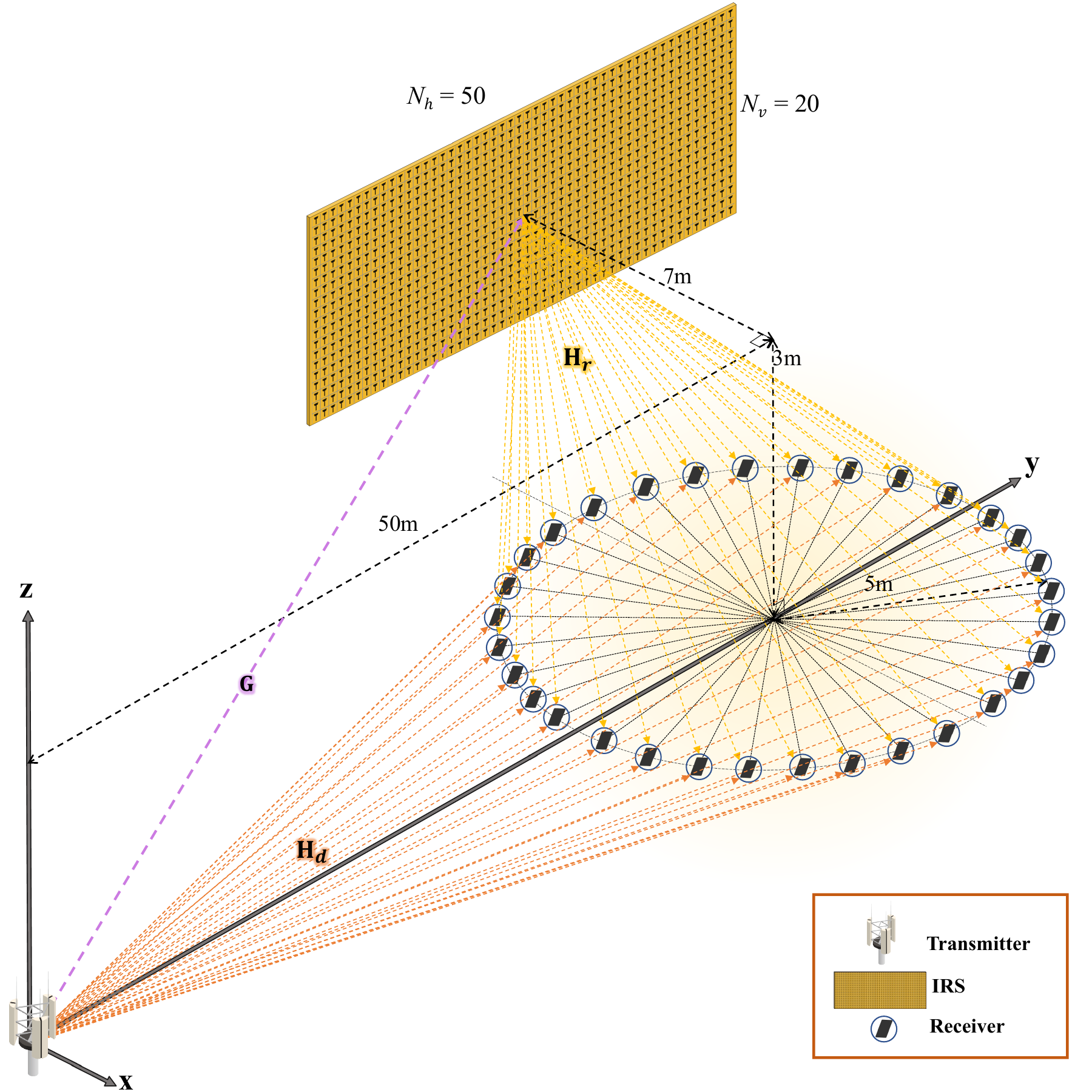}}
\vspace{6pt}
\caption{Simulated $1000$-element IRS-aided network configuration.}
\label{fig:env_setup}
\vspace{-12pt}
\end{figure}

\begin{assumption} \label{assumption: two-stage problem}
\noindent The following are in effect for problem \eqref{eqn: two-stage problem}: 
\begin{itemize}
    \item  There exists a uniform constant $C_H \geq C(\bm{H}(\bm{\theta},\omega))$ for any $\theta \in \Theta$ and a.e. $\omega \in \Omega$, where $C(\bm{H}(\bm{\theta},\omega))$ appears in \eqref{eqn: Lojasiewicz} ;
    \item For a.e. $\omega \in \Omega$, $\bm{H}(\cdot,\omega)$ is twice continuously differentiable,  uniformly bounded, Lipschitz continuous and Lipschitz smooth.
\end{itemize}
\end{assumption}
\noindent The convergence rate of the algorithm is presented next. 

\begin{theorem} \label{thm: convergence analysis}
Assume that $\{\bm{\theta}_t\}_{t = 0}^T$, $T > 0$, is generated by Algorithm \textnormal{\ref{Algorithm: iZoSGA-WMMSE}}, where $\bm{\theta}_{t^*}$ is the point that the algorithm returns. Let Assumption \textnormal{\ref{assumption: two-stage problem}} be in effect and, for each $t$, let $\varepsilon_t(\bm{\theta}_t,\omega_t) = \|\widetilde{\bm{W}}_t - \bm{W}_t^*\|$ be the oracle error at iteration $t$, where $\bm{W}_t^*$ is some solution to the second-stage problem given $\bm{\theta}_t$ and $\omega_t$. Fix some (outer) error tolerance $\epsilon > 0$. Then, by letting $T = \mathcal{O}(\sqrt{S}\epsilon^{-4})$ and $\mu = \mathcal{O}(1/\sqrt{M_U T})$, it holds that
\begin{align}
      &  \mathbb{E}\left\{\left\| \nabla \phi^{1/(2{\rho})}(\bm{\theta}_{t^*})\right\|^2\right\} =\mathcal{O}\left( \epsilon^2 + \bar{\varepsilon}\right),
\end{align}
\noindent where $\bar{\varepsilon}$ is the \emph{averaged-over-the-iterates} oracle error, defined as
\[ \bar{\varepsilon} \triangleq \frac{1}{T+1} \sum_{t=0}^T \mathbb{E}\left\{\varepsilon(\bm{\theta}_t,\omega_t)\right\}.\]
\end{theorem}
From Theorem \ref{thm: convergence analysis}, we observe that if the inexact oracle yields \eqref{eqn: second-stage problem} solutions that are, \textit{on average}, $\bar\varepsilon$-far from optimal, then applying iZoSGA algorithm propagates this error as $\sqrt{\bar\varepsilon}$ in solving the two-stage problem \eqref{eqn: two-stage problem}. In other words, iZoSGA is predictably \textit{stable} relative to inexact WMMSE-based precoding.

\begin{figure*}[t]
  \centering
  \begin{subfigure}[b]{0.33\textwidth}
  \centering
  \includegraphics[width=\textwidth, height=0.75692307692\textwidth]{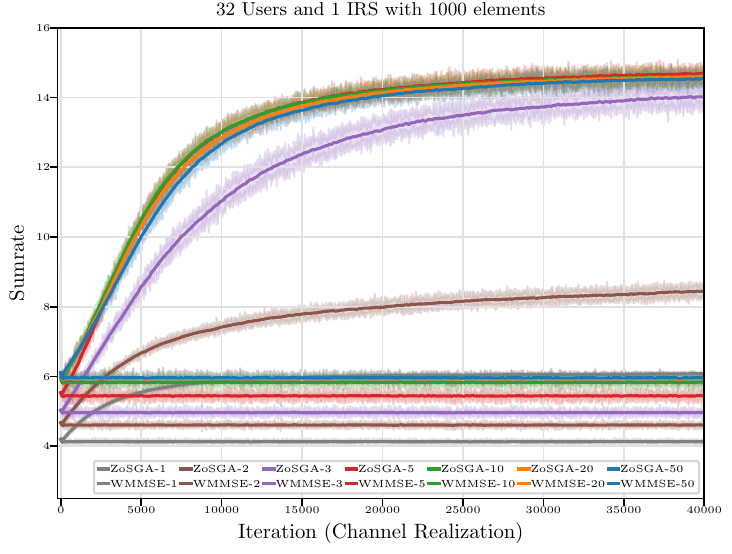}
  \caption{}
  \label{fig:1}
  \end{subfigure}
  \begin{subfigure}[b]{0.33\textwidth}
  \centering
  \includegraphics[width=\textwidth, height=0.75692307692\textwidth]{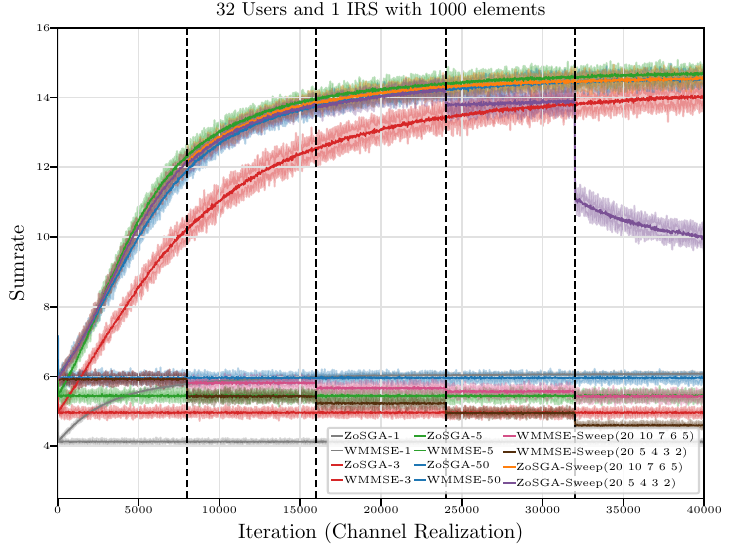}
  \caption{}
  \label{fig:2}
  \end{subfigure}
  \begin{subfigure}[b]{0.33\textwidth}
  \centering
  \includegraphics[width=\textwidth, height=0.75692307692\textwidth]{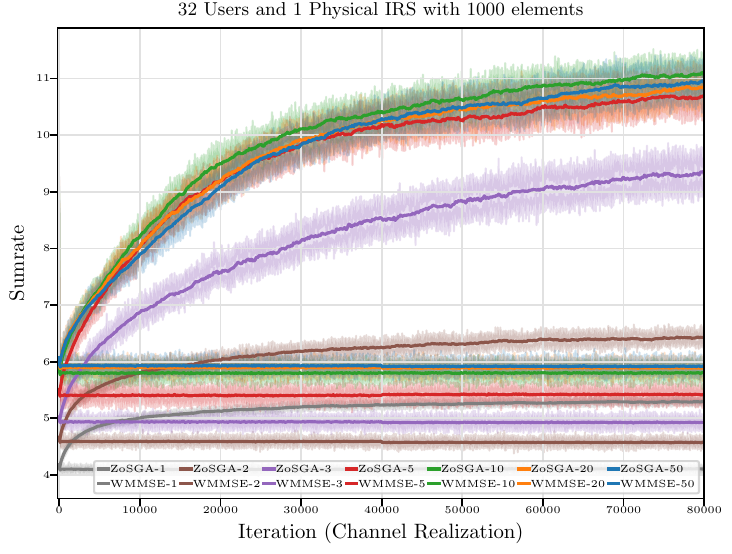}
  \caption{}
  \label{fig:3}
  \end{subfigure} 
\vspace{-10bp}
\caption{Average sumrates achieved by (a) WMMSE \cite{wmmseShi2011} (random IRS phase-shifts) and iZoSGA for 1,2,3,5,10,20,50 WMMSE iterations,  
(b) WMMSE and iZoSGA with decreasing WMMSE iterations every $8000$ steps ($20{\to}10{\to}7{\to}6{\to}5$ and $20{\to}5{\to}4{\to}3{\to}2$), and
(c) WMMSE and iZoSGA with a physical EM IRS model for 1,2,3,5,10,20,50 WMMSE iterations.
}
\label{fig:main}
\vspace{-10bp}
\end{figure*}

 Using this oracle error estimate given in \eqref{eqn: final oracle error bound} (which, in turn, is based on the natural error criterion given in \eqref{eqn: natural error criterion}) in conjunction with Theorem \ref{thm: convergence analysis}, implies that applying Algorithm \ref{Algorithm: iZoSGA-WMMSE} with tolerance $\epsilon >0$, and setting $T = \mathcal{O}(\sqrt{S}\epsilon^{-4})$, $\mu = \mathcal{O}(1/\sqrt{M_UT})$, yields
\[ \mathbb{E}\left\{\left\| \nabla \phi^{1/(2\rho)}(\bm{\theta}_{t^*})\right\|^2 \right\} = \mathcal{O}\left(\epsilon^2 + \frac{1}{T+1}\sum_{t=0}^T \mathbb{E}\left\{\widetilde{\varepsilon}(\bm{\theta}_t,\omega_t)^\xi\right\} \right).\]
\noindent This provides a theoretical grounding for the effectiveness of the subsequent numerical results presented in the next section.

\vspace{-4pt}
\section{Simulations}
\label{sec:simulations}
\vspace{-7pt}


\par As shown in Figure \ref{fig:env_setup}, we consider a simulated scenario where each receiver $k$ is connected to a common transmitter with the direct LoS link $\bm{h}_{d,k}$ as well as the indirect non-LoS link $\bm{h}_{r,k}$ directed by the IRS towards that receiver. The (total) link connecting the IRS to the transmitter is denoted by the channel matrix $\bm{G}$. We consider the Rician fading model for these intermediate channels to collectively specify a channel model for $\bm{H}(\bm{\theta},\omega)$, and also characterize the involved randomness, i.e., $\omega=\{\bm{G}, \bm{h}_{r,k},\bm{h}_{d,k}, k {\in} \mathbb{N}_K^+\}$. For clarity, Figure \ref{fig:env_setup} illustrates that $\bm{H}_d$ and $\bm{H}_r$ denote $\{\bm{h}_{d,k} {,} k {\in} \mathbb{N}_K^+\}$ and $\{\bm{h}_{r,k} {,} k {\in} \mathbb{N}_K^+\}$, respectively.
 The exact details of the channel model and pathloss gains are provided in \cite[Section V-A]{IEEE_tran_Hashmietal}.

\par In a representative configuration, we consider a transmitter with $6$ antennas (i.e., $M = 6$), $32$ receivers (i.e., $K=32$), and an IRS with $1000$ \textit{purely} phase-shifting elements denoted by $\bm{\theta} \in \mathbb{R}^S$, with $S = 1000$ (i.e., the amplitude gains of all IRS elements are fixed to unity). Under the specified network setting, the \textit{effective} channel vector $\bm{h}_k(\bm{\theta},\omega)$ from the transmitter to receiver $k$ is expressed as
\begin{align*}
    \bm{h}_k(\bm{\theta}, \omega)
    \triangleq
    \underbrace{\bm{G}^\hermtr \text{Diag}(\bm{A} \circ e^{j{\bm{\phi}}}) \bm{h}_{r,k}}_{\text{non-LOS link}} 
    + 
    \underbrace{\bm{h}_{d,k} }_{\text{LOS link}} ,
    \end{align*}
where the IRS parameters are represented by unit amplitudes $\bm{A} = \mathbf{1} \in [0,1]^{1000}$ and tunable phases $\bm{\phi} \in [-2\pi,2\pi]^{1000}$ \cite{ch_model:wu2019intelligent}. The statistics of CSI (S-CSI) of all intermediate channels in $\omega$ are assumed to be constant throughout operation of the system. Note that, for this network configuration, the total number of (cascaded) channel links is very large, specifically $38,192$ links. We emphasize at this point that the channel and specific IRS configuration are assumed here only for simulation purposes; that is, \textit{iZoSGA is completely oblivious to any channel modeling or IRS structure information}, operating only on the basis of effective channels $\bm{h}_k(\bm{\theta},\omega)$. 

As mentioned before, the short-term precoding oracle error can be controlled solely by the number of WMMSE iterations, making the empirical analysis interpretable. Finally, each experimental result is averaged over $60$ independent simulations, unless stated otherwise, showcasing average sumrates achieved by each stated method in bits per second per hertz (bps/Hz).

\par First, we compare WMMSE with unoptimized IRS parameters and fine-tuned IRS parameters using iZoSGA, for different number of WMMSE iterations, i.e., the precoders (oracle) have varying levels of \textit{inexactness}. In Figure \ref{fig:1}, we see that the reduction in $\varepsilon(\bm{\theta},\omega)$ is insignificant for $10$ or larger WMMSE iterations, and similar performance gains are achieved by iZoSGA. At 1, 2, and 3 iterations, due to higher $\varepsilon(\bm{\theta},\omega)$ magnitudes, we observe a substantial decrease in the final QoS achieved by iZoSGA. 

\par Now, an interesting observation can be made when running WMMSE for only $5$ iterations, where it shows a performance gap compared to the case of $10$ iterations. Quite remarkably, iZoSGA fine-tuning of IRS parameters closes this gap. Within 2500 (outer) iterations, iZoSGA with 5 WMMSE steps matches the performance of higher-iteration oracles, as IRS parameters compensate for precoder errors. Ultimately, iZoSGA with a 5-iteration WMMSE oracle achieves state-of-the-art performance.

\par We further evaluate the changes in $\varepsilon(\bm{\theta},\omega)$ during the learning process on the convergence of iZoSGA by performing two experiments where we change the WMMSE iterations every $8000$ outer iterations, in the order as shown in Figure \ref{fig:2}. It can be seen that the orange curve converges smoothly to the best achievable weighted sum-rate, confirming that as long as $\varepsilon(\bm{\theta},\omega)$ is below a certain threshold ($\varepsilon'$), employing iZoSGA leads to optimal performance. The purple curve shows that if, at any point, $\varepsilon(\bm{\theta},\omega) > \varepsilon'$, then we observe performance degradation of iZoSGA. In Figure \ref{fig:2}, it can be seen that the threshold lies at around 5 WMMSE iterations, and, thus, we observe a (even small) drop in the purple curve after 16000 iterations.

\par Lastly, we extend iZoSGA to a practical \textit{physical EM IRS model} based on varactor diodes \cite{phy_em_model:costa2021electromagnetic}, where each IRS parameter is a function of the corresponding tunable \textit{coupling capacitance} $C^s_{var}, s\in \mathbb{R}^+_S$ rather than directly tunable phases/amplitudes. In Figure \ref{fig:3}, we interestingly observe the same trend as before: poor performance for 1–3 WMMSE iterations, but strong performance for 5–50 WMMSE iterations. Predictably, the performance gains are lower than the ideal IRS case, since varactor constraints couple phase and amplitude, reducing the effective parameter space.

\vspace{-4pt}
\section{Conclusions} \label{sec: Conclusions}
\vspace{-7pt}
We introduced iZoSGA, a data-driven, model-free zeroth-order algorithm for two-stage stochastic IRS-assisted beamforming that combines inexact oracles for short-term precoding with zeroth-order sample gradients for IRS tuning. Our analysis, under general and realistic assumptions, establishes that iZoSGA stably converges near a stationary point of the underlying nonconvex problem, with deviation explicitly tied to the inexact oracle error. Experiments that leverage WMMSE as a surrogate oracle demonstrate that its error remains manageable, reinforcing theoretical findings and illustrating potential practical feasibility. The approach scales effectively to large multi-user networks, remains agnostic to channel, topology, or IRS structure, and is straightforward to implement; our simulations, spanning idealized to physically realistic IRS models, validate its robustness and applicability. Overall, this work provides theoretical convergence guarantees for a potentially deployable IRS tuning algorithm, offering a compelling resource allocation framework for two-stage (short/long-term) IRS-aided beamforming.

\clearpage
\bibliographystyle{IEEEbib}
\bibliography{izosga_icassp_bib}

\begin{thebibliography}{10}

\bibitem{mimo:larsson2014massive}
Erik~G Larsson, Ove Edfors, Fredrik Tufvesson, and Thomas~L Marzetta,
\newblock ``{M}assive {MIMO} for next generation wireless systems,''
\newblock {\em IEEE Commun. Mag.}, vol. 52, no. 2, pp. 186--195, 2014.

\bibitem{udn:kamel2016ultradense}
Mahmoud Kamel, Walaa Hamouda, and Amr Youssef,
\newblock ``{U}ltra-dense networks: {A} survey,''
\newblock {\em IEEE Commun. Surv. Tutor.}, vol. 18, no. 4, pp. 2522--2545, 2016.

\bibitem{udn:gotsis2016ultradense}
Antonis Gotsis, Stelios Stefanatos, and Angeliki Alexiou,
\newblock ``{U}ltra{D}ense networks: {T}he new wireless frontier for enabling 5{G} access,''
\newblock {\em IEEE Veh. Technol. Mag.}, vol. 11, no. 2, pp. 71--78, 2016.

\bibitem{5g:andrews2014will}
Jeffrey~G Andrews, Stefano Buzzi, Wan Choi, Stephen~V Hanly, Angel Lozano, Anthony~CK Soong, and Jianzhong~Charlie Zhang,
\newblock ``{W}hat will 5{G} be?,''
\newblock {\em IEEE J. Sel. Areas Commun.}, vol. 32, no. 6, pp. 1065--1082, 2014.

\bibitem{5g:shafi20175g}
Mansoor Shafi, Andreas~F Molisch, Peter~J Smith, Thomas Haustein, Peiying Zhu, Prasan De~Silva, Fredrik Tufvesson, Anass Benjebbour, and Gerhard Wunder,
\newblock ``5{G}: {A} tutorial overview of standards, trials, challenges, deployment, and practice,''
\newblock {\em IEEE J. Sel. Areas Commun.}, vol. 35, no. 6, pp. 1201--1221, 2017.

\bibitem{reactive:wu2019intelligent}
Qingqing Wu and Rui Zhang,
\newblock ``Intelligent reflecting surface enhanced wireless network via joint active and passive beamforming,''
\newblock {\em IEEE Trans. Wirel. Commun.}, vol. 18, no. 11, pp. 5394--5409, 2019.

\bibitem{reactive:wu2019towards}
Qingqing Wu and Rui Zhang,
\newblock ``{T}owards smart and reconfigurable environment: {I}ntelligent reflecting surface aided wireless network,''
\newblock {\em IEEE Commun. Mag.}, vol. 58, no. 1, pp. 106--112, 2019.

\bibitem{reactive:wang2020intelligent}
Peilan Wang, Jun Fang, Xiaojun Yuan, Zhi Chen, and Hongbin Li,
\newblock ``{I}ntelligent reflecting surface-assisted millimeter wave communications: {J}oint active and passive precoding design,''
\newblock {\em IEEE Trans. Veh. Technol.}, vol. 69, no. 12, pp. 14960--14973, 2020.

\bibitem{reactive:wu2020joint}
Qingqing Wu and Rui Zhang,
\newblock ``Joint active and passive beamforming optimization for intelligent reflecting surface assisted swipt under qos constraints,''
\newblock {\em IEEE J. Sel. Areas Commun.}, vol. 38, no. 8, pp. 1735--1748, 2020.

\bibitem{sco:yang2021}
Zhaohui Yang, Mingzhe Chen, Walid Saad, Wei Xu, Mohammad Shikh-Bahaei, H~Vincent Poor, and Shuguang Cui,
\newblock ``{E}nergy-efficient wireless communications with distributed reconfigurable intelligent surfaces,''
\newblock {\em IEEE Trans. Wirel. Commun.}, vol. 21, no. 1, pp. 665--679, 2021.

\bibitem{10123709}
Jianzheng Li, Weijiang Wang, Rongkun Jiang, Xinyi Wang, Zesong Fei, Shihan Huang, and Xiangnan Li,
\newblock ``{P}iecewise-{DRL}: {J}oint beamforming optimization for {RIS}-assisted {MU}-{MISO} communication system,''
\newblock {\em IEEE Internet Things J.}, vol. 10, no. 19, pp. 17323--17337, 2023.

\bibitem{ml:taha2021}
Abdelrahman Taha, Muhammad Alrabeiah, and Ahmed Alkhateeb,
\newblock ``{E}nabling large intelligent surfaces with compressive sensing and deep learning,''
\newblock {\em IEEE Access}, vol. 9, pp. 44304--44321, 2021.

\bibitem{ml:yang2021}
Bo~Yang, Xuelin Cao, Chongwen Huang, Chau Yuen, Lijun Qian, and Marco Di~Renzo,
\newblock ``{I}ntelligent spectrum learning for wireless networks with reconfigurable intelligent surfaces,''
\newblock {\em IEEE Trans. Veh. Technol.}, vol. 70, no. 4, pp. 3920--3925, 2021.

\bibitem{drl:mismar2019dqn}
Faris~B Mismar, Brian~L Evans, and Ahmed Alkhateeb,
\newblock ``{D}eep reinforcement learning for {5G} networks: Joint beamforming, power control, and interference coordination,''
\newblock {\em IEEE Trans. Commun.}, vol. 68, no. 3, pp. 1581--1592, 2019.

\bibitem{drl:huang2020ddpg}
Chongwen Huang, Ronghong Mo, and Chau Yuen,
\newblock ``{R}econfigurable intelligent surface assisted multiuser miso systems exploiting deep reinforcement learning,''
\newblock {\em IEEE J. Sel. Areas Commun.}, vol. 38, no. 8, pp. 1839--1850, 2020.

\bibitem{SSCO_1}
An~Liu, Vincent K.~N. Lau, and Min-Jian Zhao,
\newblock ``Online successive convex approximation for two-stage stochastic nonconvex optimization,''
\newblock {\em IEEE Trans. Signal Process.}, vol. 66, no. 22, pp. 5941--5955, 2018.

\bibitem{SSCO_2}
An~Liu, Vincent K.~N. Lau, and Borna Kananian,
\newblock ``{S}tochastic successive convex approximation for non-convex constrained stochastic optimization,''
\newblock {\em IEEE Trans. Signal Process.}, vol. 67, no. 16, pp. 4189--4203, 2019.

\bibitem{sco:zhao2020tts}
Ming-Min Zhao, Qingqing Wu, Min-Jian Zhao, and Rui Zhang,
\newblock ``{I}ntelligent reflecting surface enhanced wireless networks: {T}wo-timescale beamforming optimization,''
\newblock {\em IEEE Trans. Wirel. Commun.}, vol. 20, no. 1, pp. 2--17, 2020.

\bibitem{IEEE_tran_Hashmietal}
H.~Hashmi, S.~Pougkakiotis, and D.~Kalogerias,
\newblock ``{M}odel-free learning of two-stage beamformers for passive {IRS}-aided network design,''
\newblock {\em IEEE Trans. Signal Process.}, vol. 72, pp. 652--669, 2024.

\bibitem{wmmseShi2011}
Qingjiang Shi, Meisam Razaviyayn, Zhi~Quan Luo, and Chen He,
\newblock ``{A}n iteratively weighted {MMSE} approach to distributed sum-utility maximization for a {MIMO} interfering broadcast channel,''
\newblock {\em IEEE Trans. Signal Process.}, vol. 59, pp. 4331--4340, 2011.

\bibitem{LojasiewiczInequality}
S.~\L{}ojasiewicz,
\newblock ``{S}ur la g\'eom\'etrie semi- et sous- analytique,''
\newblock {\em Ann. Inst. Fourier}, pp. 1575--1595, 1993.

\bibitem{SIAMOpt:Davis}
Damek Davis and Dmitriy Drusvyatskiy,
\newblock ``{S}tocahstic model-based minimization of weakly convex functions,''
\newblock {\em SIAM J. Optim.}, vol. 29, no. 1, pp. 207--239, 2019.

\bibitem{ch_model:wu2019intelligent}
Qingqing Wu and Rui Zhang,
\newblock ``{I}ntelligent reflecting surface enhanced wireless network via joint active and passive beamforming,''
\newblock {\em IEEE Trans. Wirel. Commun.}, vol. 18, no. 11, pp. 5394--5409, 2019.

\bibitem{phy_em_model:costa2021electromagnetic}
Filippo Costa and Michele Borgese,
\newblock ``{E}lectromagnetic model of reflective intelligent surfaces,''
\newblock {\em IEEE Open J. Commun. Soc.}, vol. 2, pp. 1577--1589, 2021.

\end{thebibliography}

\end{document}